\RequirePackage{fix-cm}
\PassOptionsToPackage{dvipsnames}{xcolor}

\documentclass{article}

\usepackage[nonatbib, preprint]{neurips_2024}
\usepackage[numbers]{natbib}

\usepackage{multicol}
\usepackage{multirow}
\usepackage{xspace}
\usepackage{setspace}
\usepackage{listings}
\usepackage{microtype}
\usepackage{tcolorbox}
\usepackage{hyperref}
\usepackage{ifthen}
\usepackage{graphicx}
\usepackage{tabularx}
\usepackage{ragged2e}
\usepackage{wrapfig}
\usepackage{amsmath}
\usepackage{amssymb}

\hypersetup{colorlinks= true,
            bookmarksnumbered = true, 
            citecolor = {PineGreen}, 
            urlcolor = {blue},		
            filecolor = {blue}, 
            linkcolor = {blue}}
\usepackage[nameinlink]{cleveref}
\hypersetup{bookmarksdepth=-2}

\crefname{algocf}{alg.}{algs.}
\Crefname{algocf}{Algorithm}{Algorithms}

\usepackage{siunitx}

\usepackage[caption=false,font=footnotesize,labelfont=sf,textfont=sf]{subfig}

\usepackage[export]{adjustbox}
\usepackage[utf8x]{inputenc}

\newboolean{showcomments}
\setboolean{showcomments}{true} % comment this line to deactivate comments
\ifthenelse{\boolean{showcomments}}{
  \newcommand{\nbc}[3]{
    {\colorbox{#3}{\bfseries\sffamily\scriptsize\textcolor{white}{#1}}}%
    {\textcolor{#3}{\textsf\small$\blacktriangleright$\textit{#2}$\blacktriangleleft$}}}
  \newcommand{\todo}[1]{\nbc{TODO}{#1}{blue}\xspace}
}{
  \newcommand{\nbc}[3]{}
  \renewcommand{\todo}[1]{}
}

\newcommand{\etal}{\hbox{\emph{et al.}}\xspace}

\newcommand{\ie}{\hbox{\emph{i.e.}}\xspace}
\newcommand{\etc}{\hbox{\emph{etc.}}\xspace}
\newcommand{\vs}{\hbox{\emph{vs}}\xspace} 

\DeclareFixedFont{\ttb}{T1}{txtt}{bx}{n}{8} % for bold
\DeclareFixedFont{\ttm}{T1}{txtt}{m}{n}{8}  % for normal

\usepackage{color}
\definecolor{deepblue}{rgb}{0,0,0.5}
\definecolor{deepred}{rgb}{0.6,0,0}
\definecolor{deepgreen}{rgb}{0,0.5,0}

\usepackage{listings}

\newcommand\pythonstyle{\lstset{
language=Python,
basicstyle=\small\ttm,
otherkeywords={self},             % Add keywords here
keywordstyle=\ttb\color{deepblue},
emphstyle=\ttb\color{deepred},    % Custom highlighting style
stringstyle=\color{deepgreen},
showstringspaces=false            % 
}}

\lstnewenvironment{python}[1][]
{
\pythonstyle
\lstset{#1}
}
{}

\newcommand\pythoninline[1]{{\pythonstyle\lstinline!#1!}}
\usepackage{pgfplotstable}
\pgfplotsset{compat=1.17}

\usepackage{combelow}

\title{Bringing Structure to Naturalness:\\ On the Naturalness of ASTs}

\author{%
  Profir-Petru~P\^arțachi\\
  National Institute of Informatics\\
  Tokyo, Japan\\
  \texttt{profir@nii.ac.jp}\\
  \And
  Mahito~Sugiyama\\
  National Institute of Informatics\\
  Tokyo, Japan\\
  The Graduate University for Advanced Studies, SOKENDAI\\
  Kanagawa, Japan\\
  \texttt{mahito@nii.ac.jp} \\
}

\begin{document}
\maketitle

\begin{abstract}
    Source code comes in different shapes and forms. 
Previous research has already shown code to be more predictable than natural language as well as highlighted its statistical predictability at the token level: 
source code can be natural. 
More recently, the structure of code --- 
control flow, syntax graphs, abstract syntax trees \etc --- 
has been successfully used to improve the state-of-the-art on numerous tasks: 
code suggestion, code summarisation, method naming \etc 
This body of work implicitly assumes that structured representations of code are similarly statistically predictable, 
\ie that a structured view of code is also natural. 
We consider that this view should be made explicit and propose directly studying the Structured Naturalness Hypothesis. 
Beyond just naming existing research that assumes this hypothesis and formulating it, 
we also provide evidence in the case of trees: 
TreeLSTM models over ASTs for some languages, such as Ruby, are competitive with $n$-gram models while handling the syntax token issue highlighted by previous research `for free'.
For other languages, such as Java or Python, we find tree models to perform worse, suggesting that downstream task improvement is uncorrelated to the language modelling task.
Further, we show how such naturalness signals can be employed for near state-of-the-art results on just-in-time defect prediction while forgoing manual feature engineering work.

\end{abstract}

\title{Bringing Structure to Naturalness: On the Naturalness of ASTs}

\section{Introduction}
\label{sec:introduction}

Source code is highly repetitive; despite the flexibility that programming languages offer, developers tend to converge to similar constructions which lend to its statistical predictability~\cite{hindle2012naturalness}. 
In the seminal work on source-code \emph{naturalness}, Hindle \etal~\cite{hindle2012naturalness} show that this predictability, which represents naturalness, lends itself well to statistical language models. 
This property of source-code has been successfully exploited for many tasks such as code synthesis~\cite{allamanis2015bimodal, gvero2015synthesizing}, code completion~\cite{bielik2016phog, raychev2016learning}, identifier or method name suggestion~\cite{refinym2018dash, allamanis2016convolutional}, defect prediction~\cite{wang2016automatically} \etc 
Allamanis \etal provide a more complete overview in their survey~\cite{allamanis2018survey}.
More recently, Rahman \etal~\cite{rahman2019natural} show that much of this predictability is due to syntax tokens and removing these, akin to removing stop-words, makes code less predictable than previously claimed. 
In addition to this, Rahman \etal consider a graph view that is obtained from a program-dependency graph which does not have to contend with syntax tokens by mining (2,3,4)-hop sub-graphs. 
They show that this view is more predictable than a token view; however, they stop short of formulating structured naturalness or studying other structures more extensively.

Recent work has started exploiting structure, not necessarily restricted to the view presented by Rahman \etal, either as additional signal~\cite{liu2016neural, shido2019automatic, yin2017syntactic, alon2018code2seq}, or to overcome limitations of existing methods: such as overcoming vanishing gradients in neural models using a graph structure superimposed on the tokens~\cite{allamanis2017learning} by providing shorter paths for information to propagate along.
While it seems intuitive that naturalness should extend from the unstructured token domain into structures that we obtain, which we refer to as \emph{structured naturalness}, either via parsing, such as abstract syntax trees (ASTs), or by static or dynamic analysis, such as control-flow graphs, data-flow graphs, call graphs \etc, this assumption has not been studied directly.
Further, it is an open question if such tasks help with the language modelling task or specifically the downstream tasks for which such views are employed.

We propose that the structured naturalness hypothesis should be the primary subject of study rather than an unstated assumption as the current body of literature would have it. 
By better knowing the nature of the statistical regularities induced in structured views of code from the underlying naturalness of the source code itself, we can better tune statistical approaches that seek to use it either as an aid or directly as a signal.
We also wish to explore if the structure is task-specific, and if it should be considered as a prior or inductive bias.

Rahman \etal has placed the first stones towards the study of structured naturalness by looking at a mined sub-graph view of code; however, that mimics $n$-grams too closely and misses the more fundamental nature of structured naturalness by casting it too close to the token-level view.  
In this work, we propose, first of all, to formulate the hypothesis and, then, to study the naturalness of ASTs directly to demonstrate the promises of structured naturalness. 
Existing work already makes use of tree structures for code~\cite{liu2016neural, shido2019automatic, yin2017syntactic, allamanis2017learning}, hence we aim to retroactively provide empirical evidence for why such language models should work.
Instead of following too closely the original Hindle \etal~\cite{hindle2012naturalness} naturalness work and restricting ourselves to a $n$-gram analogue, as there is no canonical language model, we instead consider TreeLSTM with masked token prediction as the training task.
Our choice of using ASTs is both pragmatic --- it allows us to use existing parsers --- and as a representative structure that we expect to exhibit structured naturalness that has not been previously studied as a first-class citizen of such research.
We find mixed results across the languages, suggesting that perhaps downstream tasks benefit from the structure as a side-channel rather than enhanced language modelling directly.
To explore if structure helps downstream tasks specifically, rather than by improving language understanding, we explore if the statistical signal can be employed by a simple classification pipeline trained to perform just-in-time defect prediction.
We find that even a simple pipeline can reach near state-of-the-art results on JIT defect prediction with no manual input for feature selection.

Structured naturalness holds promises beyond the domain of programming languages: going from the naturalness of tokens to that of parses can extend to natural languages.
We can already see the idea of employing graph structures for summarisation in work from Allamanis \etal~\cite{allamanis2017learning}.
They endow an English sentence with either entity recognition-based graphs or dependency parse graphs to avoid vanishing gradient issues.
While they use this construction for pragmatic reasons, this still leaves the door open to questions about the regularities of such parses or graphs, hence their structured naturalness.
Further, there are open questions about representations that can be antagonistic towards pre-training tasks (masked token prediction) while being synergistic to downstream tasks\footnote{We make our experimentation code available here: \url{https://drive.proton.me/urls/07K7T7ABEW\#kk74oCX54fUk}}.

To summarise, we:
\begin{itemize}
    \item propose to extend the naturalness hypothesis to structured representations of code: both from parses, such as ASTs, and from analysis, such as control-flow graphs, call graphs \etc
    \item study the naturalness of AST trees, providing empirical evidence to justify why such models should succeed at source-code prediction tasks or, indeed, why they do not.
    \item demonstrate near state-of-the-art results for just-in-time prediction without requiring manual feature engineering work to show how regularities from ASTs can help downstream tasks.
\end{itemize}

\section{Structured Naturalness}
\label{sec:problem_statement}

\begin{figure}[!t]
    \centering
    \includegraphics[width=0.66\textwidth]{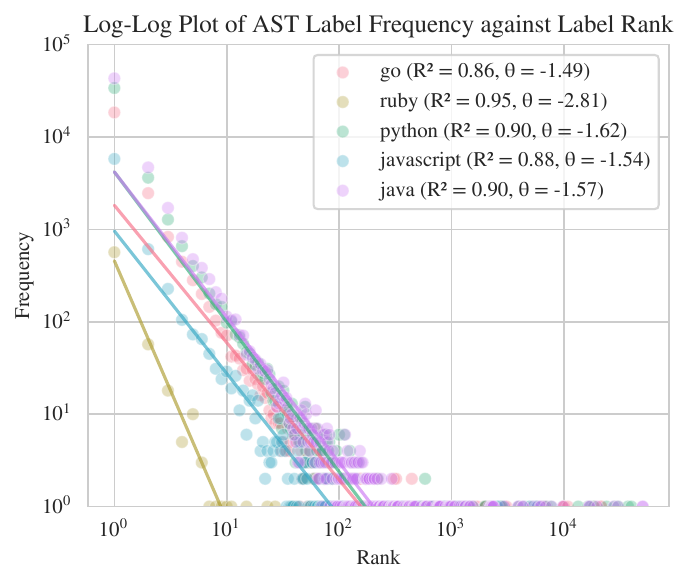}
    \caption[Log-Log Plot of AST Label Frequency against Rank]{Log-Log plot of AST Labels (inner nodes) against their rank by frequency in the corpus. The plots show clear linear trends in log scale across the languages indicative of Zipf's law (negative slopes ($\theta$) and r$^2$-values above $0.86$). Thus we expect a similar naturalness in tree contexts as was previously shown in $n$-gram models at the token level.}
    \label{fig:zipf}
\end{figure}

In their seminal work on naturalness, Hindle \etal~\cite{hindle2012naturalness} show that source code has statistical regularities and, further, how language models can exploit these for a variety of tasks, including those that require generative models, discriminative \etc
It is intuitive to assume that these regularities, of a Zipfian nature, at the token level extend to structures we can derive from source code: abstract syntax trees, control flow graphs, data-flow graphs, program dependency graphs, call graphs \etc
Existing literature assumes this either implicitly~\cite{liu2016neural, shido2019automatic, yin2017syntactic, alon2018code2seq} by employing models that would require such regularities to exist for the learning process (TreeLSTMs~\cite{tai2015improved}, TreeTransformers~\cite{wang2019tree} as well as ASTTransformers~\cite{tang2021ast}, or linearisations of ASTs as sets of AST paths modelled via seq2seq models~\cite{alon2018code2seq}) or makes an explicit, yet pragmatic, choice for priors: such as code idiom work assuming a Pitman-Yor prior~\cite{allamanis2014mining}, which represents the same Zipfian nature Hindle \etal observed extended to tree-like structures.
More recent work exposes structure selectively, after linearisation, to large language models as in-context or few-shot learning during the prompting process~\cite{ahmed2023improving}.
To the best of our knowledge, there are no examples where generative models are employed for semantic (graph) structures of code;
however, we can find examples that exploit syntactic structure (such as ASTs) for generative processes, in no small part due to natural starting points --- often only one choice --- \ie the root of an AST. 

Further, predictive models can be pragmatically useful for practitioners; however, our wish for generative models is driven by our goal of explaining the statistical phenomenon that arises from our collective development efforts.

To further our intuition and justify empirically why ASTs should be considered ``natural'', we also looked at AST label distribution, i.e. tokens that could occur under a root-path encoding of trees. We found that such AST labels (even when ignoring leaves) show a clear Zipfian trend (\Cref{fig:zipf}).
If we ignore the long tail of single occurrences, we find that the r$^2$-values using ordinary least squares for all languages is above $0.86$.
We hypothesise that the Zipfian nature of code extends not only to syntactic structures (ASTs) but also to semantic structures (graphs).
Under fixed tooling (compiler, linker) and environment (OS, target architecture), we expect both syntactic views of code (ASTs) and semantic views of code: control-flow graphs, program dependency graphs, call graphs \etc, to be the same under fixed tooling for the same set of tokens. 

From a syntactic view, we expect the Zipfian nature to extend directly.
To observe this, consider the following.
From Hindel \etal~\cite{hindle2012naturalness} we already expect a Zipfian distribution over $n$-grams.
For each $n$-gram, we can build an associated AST fragment by considering the sub-tree rooted at the lowest common ancestor for all leaves contained in the $n$-gram.
As explained so far, we can expect the Zipfian nature to extend to AST fragments; however, there is an aspect we have to account for first: the sub-tree could contain leaves from outside the $n$-gram leaving us with the choice to either remove the paths from the root to leaves outside the $n$-gram potentially creating a nonsensical AST or drop such ASTs from the data.
The first option maintains the Zipfian nature but might not be pragmatically useful, while the latter could change the distribution and should be studied empirically.

Meanwhile, from a semantic view, we argue that developers want to avoid mentally executing code, a costly and error-prone exercise.
This leads to the assumption that identical or similar code should lead to identical or similar behaviour.
A well-understood code construct is more likely to be used leading to a rich-get-richer effect.
Hence, this likely induces a Zipfian distribution over sub-graphs in semantic views of code.

Often, obvious facts are the most difficult to prove. 
The literature already successfully exploits syntactical structure with success by making pragmatic choices.
However, exploring the statistical regularities of code structure holds the promise of more appropriate and direct statistical tooling.
A better understanding of the statistical distributions behind such structures holds the promise to improve our choices of ---
\emph{priors}: if we know from what family of distributions we are drawing samples, we can bake this assumption explicitly in the model reducing variance at the cost of bias. Further, this would enable other researchers to adjust this prior when adapting the method to new settings; 
\emph{features}: knowing the shape and regularities of our data can allow us to do feature engineering more effectively than codifying human intuition; or 
\emph{neural architecture}: as with priors, knowing the shape of our data can allow us to pick correct inductive biases and reduce the data cost of neural approaches for SE where data scarcity can be an issue;
\emph{prompt engineering}: knowing the regularities from software can aid in the selection of data to add to a prompt for in-context learning and to help a large language model provide better output for source-code manipulation tasks.
\section{Methodology}
\label{sec:method}

Previous work looking both at the naturalness of software at the token level~\cite{hindle2012naturalness, rahman2019natural} and even higher levels such as (sub-)graphs~\cite{rahman2019natural} has been biased by an $n$-gram view. 
At the token level, this has no major consequence since the $n$-gram model is a ``natural'' choice; 
at the graph or tree level, other views should be considered in the interest of ``naturalness''. 
Towards this goal, we look to existing software engineering literature and how they exploit structure to improve predictive models: GNNs over control flow graphs or syntax graphs, TreeLSTMs~\cite{tai2015improved, shido2019automatic} or TreeTransformers~\cite{wang2019tree, tang2021ast} over AST views. In this study, we focus on TreeLSTMs as an avatar for the tree model. Prior to this work, tree models have not been actively and directly researched from a ``naturalness'' perspective.
Further, LSTMs~\cite{hochreiter1997long} have long been a strong model in software engineering, and generalising to TreeLSTMs incorporates structure. 
While Transformer models are popular now in the age of large language models, their computational cost makes them prohibitive for our study where models need to be trained multiple times for cross-fold validation.

As we wish to analyse a tree model based on its entropy, firstly, we should make clear which probability concretely we are modelling:
\[
P(t_i \mid t_1, \dots, t_{i-1}, t_{i+1}, \dots, t_n, \Delta_{\top} \setminus \Delta_{\text{LCA}(t_i)}),
\]
where $t_i$ represents the $i$-th token, $\Delta_{\cdot}$ represents the (sub-)tree rooted at $\cdot$, $\text{LCA}(t_i)$ represents the first ancestor of the leaf $t_i$ that has multiple children in the AST, and we use `$\setminus$' to denote removing a sub-tree from a tree that contains it. 
We consider this probability to be an analogue for the masked token training typical of many language models, for example in the pre-training of transformers.
For convenience, we will refer to this probability as:
\[
P(t_i \mid \mathcal{K}_i).
\]
where $\mathcal{K}_i$ represents the bi-lateral and tree context.

\begin{figure}[t]
\centering
\subfloat[Example AST Fragment derived from a Python function declaration.]{\includegraphics[width=0.49\textwidth,
trim=0 0 0 0]{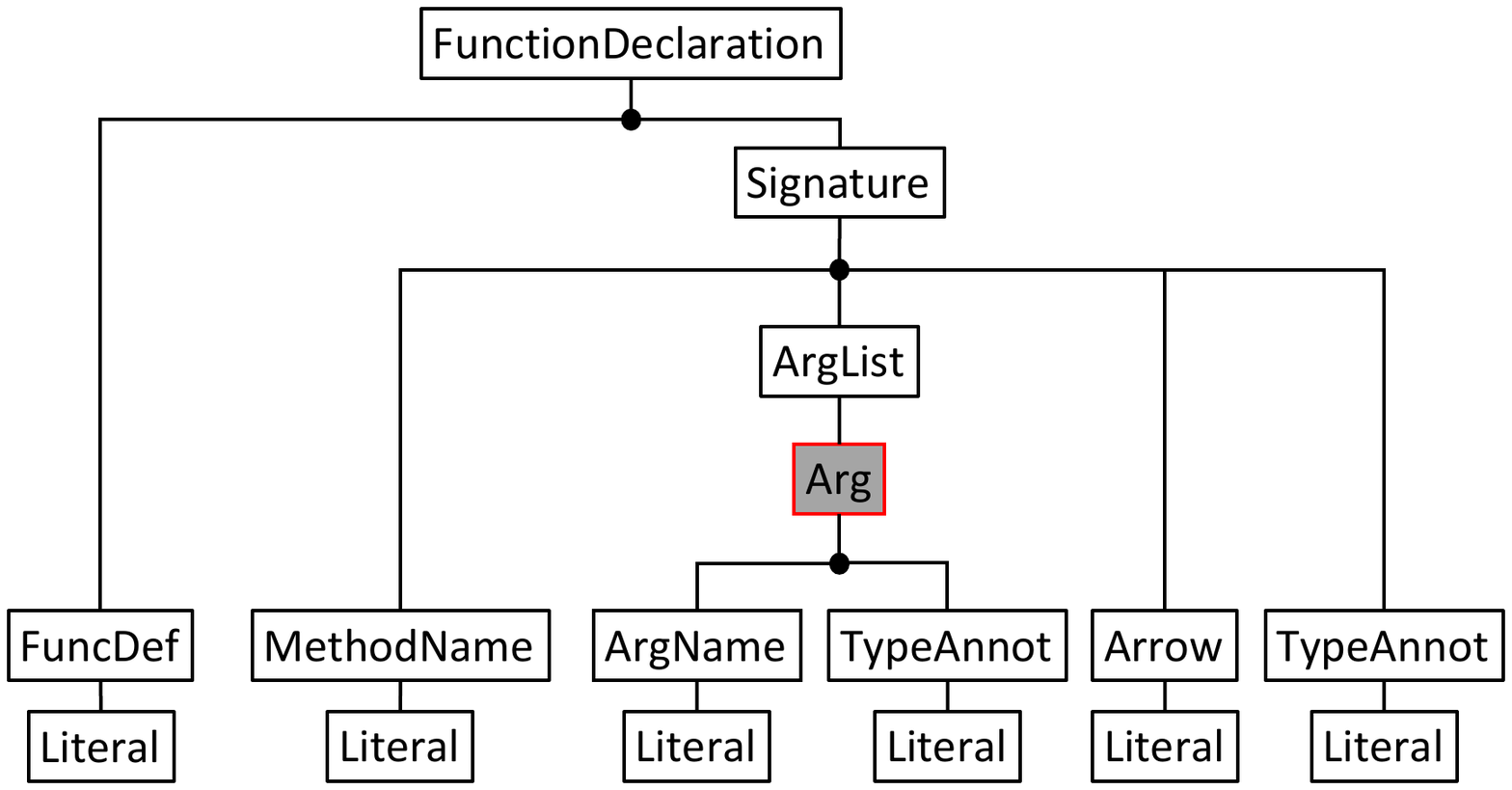}%
\label{fig:method:ast_ex}}
\hfill
\centering
\subfloat[The TreeLSTM cell corresponding to the highlighted node in \Cref{fig:method:ast_ex}.]{\includegraphics[width=0.49\textwidth, trim=0 0 0 0]{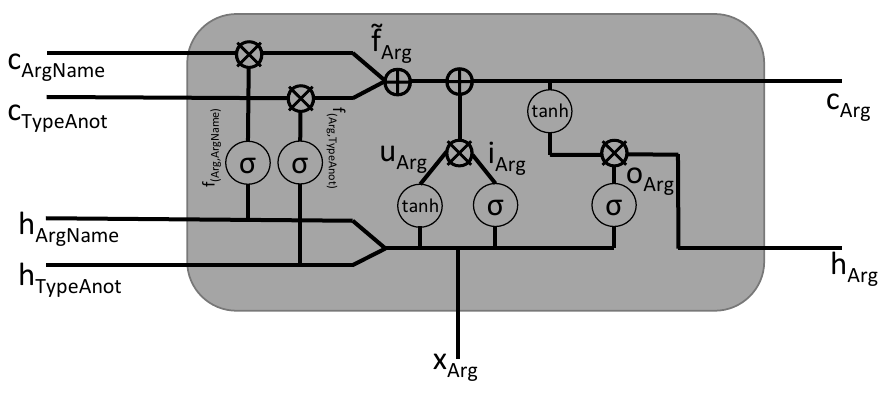}%
\label{fig:method:treelstmcell}}
\hfill
\caption{\Cref{fig:method:ast_ex} presents an example AST fragment. \Cref{fig:method:treelstmcell} presents the TreeLSTM cell for the highlighted ``Arg'' node in \Cref{fig:method:ast_ex}. We annotate the paths in \Cref{fig:method:treelstmcell} with the corresponding labels from Equations (1)--(7) replacing the subscripts with the corresponding node labels from the AST fragment.}
\label{fig:method:ast_and_cell}
\end{figure}

To estimate this probability, we make use of TreeLSTMs.
These models augment the seq2seq models to consider ancestor and sibling relationships, specifically: TreeLSTMs use a Child-Sum summarisation step to incorporate sibling relationships:
\begin{align}
\tilde{h}_j &= \sum_{k \in C(j)} h_k;\\
i_j &= \sigma\left( \mathbf{W}^{(i)}x_j + \mathbf{U}^{(i)}\tilde{h}_j+b^{(i)} \right);\\
f_{jk} &= \sigma \left( \mathbf{W}^{(f)}x_j + \mathbf{U}^{(f)}h_k+b^{(f)} \right) \text{ for } k \in C(j);\\
o_j &= \sigma\left( \mathbf{W}^{(o)}x_j + \mathbf{U}^{(o)}\tilde{h}_j+b^{(o)} \right);\\
u_j &= \tanh{\left( \mathbf{W}^{(i)}x_j + \mathbf{U}^{(i)}\tilde{h}_j+b^{(i)} \right)};\\
c_j &= i_j \odot u_j + \sum_{k \in C(j)} f_{jk} \odot c_k;\\
h_j &= o_j \odot \tanh(c_j);
\label{eq:final_lstm_state}
\end{align}
where $h_j$ is the hidden state of the $j$-th tree node,
$x_j$ is the input/feature vector of the $j$-th tree node,
$C(j)$ is the set of children of the $j$-th node,
$\mathbf{W}^{\cdot}$ are learnable parameters over input data,
$\mathbf{U}^{\cdot}$ are learnable parameters over the children hidden states,
$b^{\cdot}$ are learnable bias terms,
and $\odot$ is point-wise multiplication.
Importantly, the base case of the TreeLSTM happens at the leaves, where $C(j)$ is the empty set, hence $\tilde{h}_j$ is a zero-vector and the other features are only defined in terms of $x_j$. \Cref{fig:method:ast_and_cell} presents an AST fragment together with an annotated TreeLSMT cell for the equations above.

Finally, we predict tokens using the hidden state for each token via a simple feed-forward network that takes the final hidden state as input and outputs a token prediction, \ie:
\begin{align}
P(t_i \mid \mathcal{K}_i)\ \tilde{=}\ P(t_i \mid h_i),    
\end{align}
where $h_i$ is calculated according to \Cref{eq:final_lstm_state}.

Crucially, these models are used in an auto-regressive fashion and trained using masked token prediction to ensure that they learn the statistical patterns present in AST data sets.
Another core point is that while we use tree input, we make token-level predictions, thus our models are tree2seq.

Given a model that approximates our desired distribution, we can now focus on ``naturalness''. 
Specifically, we can look at the cross-entropy of the model when applied to a held-out set of examples. 
To approximate this, borrowing from Hindle \etal~\cite{hindle2012naturalness}, we employ cross-fold validation, and, similarly, call the resulting average across cross-folds self-cross-entropy.
Unlike Hindle \etal, we restrict ourselves to fewer folds due to the computational cost of neural models: we split the data into five groups, train on four of these and evaluate on the held-out set. 
In our preprocessing, we first determine a token that is to be masked, together with the associated sub-tree that could induce data leakage. 
Hence, for each tree, we have one prediction target, a uniformly-at-random selected token.
Concretely, when combined with cross-fold validation, what we are measuring is:
\begin{align}
    H(D,M) = -\frac{1}{n}\sum_{i=1}^n\log_2P(t_{f(i)} \mid h_{f(i)}),
\end{align}
where data-set $D$ has $n$ examples, $f$ maps document index ($i$) to masked token index that should be predicted, and model $M$ determines $h_{f(i)}$.
The number can be interpreted as a measure of how many bits of information about the data set the model has managed to capture.

In the original paper, Hindle \etal~\cite{hindle2012naturalness} consider an $n$-gram language modelling task where the task is next-word prediction. 
Different from them, as we do not employ $n$-grams, we do not have an $n$ to vary as we explore the predictability of data sets under tree models. 
These models do not have a natural connection to the $n$ in the $n$-gram model employed by the original naturalness study, 
studying them can still reveal how predictable source code is under a structured view. 
Instead, as the tree model considered generalises seq2seq models, and due to computational constraints, we place a limit on the maximum sequence length and vary that as a proxy for the $n$ in $n$-gram models.
We further diverge due to the use of our bilateral and tree contexts which can differ from $n$-grams.

We choose hyper-parameters from literature instead of performing hyper-parameter optimisation as the aim of our study is to show the self-cross-entropy of a reasonable baseline model that employs tree structure rather than optimising for, or, indeed, over-fitting to a particular language modelling task.
TreeLSTMs as used for source code employed the hyper-parameter defaults from NLP literature~\cite{shido2019automatic}; therefore we do the same. 
Specifically, our hyper-parameters are: 
$150$ for the embedding size, 
$75$ for the LSTM memory cell size, 
and $25$ for the final feed-forward network performing token prediction. 
We use a batch size of $64$ (for go, Java, and Python, due to VRAM constraints, we lower the batch size to $8$), 
a learning rate of $0.025$, 
and a weight decay of $0.0001$ together with an ADAM optimiser~\cite{kingma2014adam}.
The model is implemented in PyTorch~\cite{wallach2019pytorch}.
\section{Naturalness under TreeLSTMs}
\label{sec:results}

Previous work has shown that source code is as predictable as English~\cite{hindle2012naturalness, rahman2019natural} or even more predictable when a graphlet-like-view is considered~\cite{rahman2019natural}.
Existing literature already works with implied assumptions that ASTs are similarly `natural'. 
In this work, we want to test the hypothesis directly.
We find that ASTs are indeed natural when appropriately constructed and tackle previous challenges from a token view `for free'; however, for some languages, TreeLSTMs display less statistical predictability (higher entropy) despite being successfully applied to downstream tasks.

In the remainder of the section, first, we detail how we build an AST corpus starting from CodeSearchNet~\cite{Husain2019}, then we discuss the results of five-fold cross-validation, and finally, we compare with previous work by Rahman \etal~\cite{rahman2019natural}. 
Crucially, the AST view handles the syntax tokens for free avoiding the preprocessing needed for a token view and previously highlighted by Rahman \etal

\subsection{Corpus}
\label{sec:results:corpus}

\begin{table*}[t]
    \centering
    \caption[Corpus Statistics]{Starting from the CodeSearchNet Corpus~\cite{Husain2019} we preprocess it with ANTLR4~\cite{antlr4} to generate ASTs for each snippet. In this table, we show the statistics of the resulting corpus of ASTs and code snippets.}
    \label{tab:results:corpora}
    {\footnotesize
\begin{tabularx}{\textwidth}{ l X X X X X X X X X X X } 
\hline 
\multicolumn{1}{c}{\textbf{Lang}} & \multicolumn{1}{c}{\#} & \multicolumn{1}{c}{Voc.} & \multicolumn{3}{c}{Tree Size} & \multicolumn{3}{c}{Tree Depth} & \multicolumn{3}{c}{Sequence Size} \\
\cline{4-12}
& & & \multicolumn{1}{c}{Min} & \multicolumn{1}{c}{Med} & \multicolumn{1}{c}{Max} & \multicolumn{1}{c}{Min} & \multicolumn{1}{c}{Med} & \multicolumn{1}{c}{Max} & \multicolumn{1}{c}{Min} & \multicolumn{1}{c}{Med} & \multicolumn{1}{c}{Max}\\
\hline 
go & \multicolumn{1}{r}{234890} & \multicolumn{1}{r}{398371} & 44 & 149.5 & 396303 & 7 & 18.0 & 124 & 20 & 52.0 & 130888\\
%php & \multicolumn{1}{r}{44572} & \multicolumn{1}{r}{4555894} & 10 & 40.0 & 18025 & 7 & 7.0 & 80 & 2 & 16.5 & 8960\\
ruby & \multicolumn{1}{r}{1865} & \multicolumn{1}{r}{8517} & 22 & 140.0 & 1143 & 9 & 34.0 & 113 & 7 & 31.0 & 283\\
py & \multicolumn{1}{r}{433184} & \multicolumn{1}{r}{2649581} & 64 & 490.0 & 151608 & 25 & 59.0 & 253 & 26 & 88.0 & 28488\\
js & \multicolumn{1}{r}{63386} & \multicolumn{1}{r}{213481} & 45 & 199.0 & 1581457 & 12 & 25.0 & 419 & 24 & 87.0 & 630581\\
java & \multicolumn{1}{r}{469219} & \multicolumn{1}{r}{1077875} & 55 & 328.0 & 477296 & 10 & 55.0 & 979 & 20 & 65.0 & 68278\\
\hline
\end{tabularx} 
}

\end{table*}

To construct our corpus, we start from the public corpus provided by Husain \etal --- CodeSearchNet~\cite{Husain2019}.
CodeSearchNet represents a data set across 6 languages --- Go, Java, JavaScript, PHP, Python, and Ruby --- that pairs documentation strings with methods to which these are related.
It originally served as a parallel corpus for code retrieval, however, it has found use beyond the original task.
While the corpus does provide a pre-tokenised data set, it is not useful for our AST study.

In our circumstance, we need to further process the snippets provided to convert them to ASTs.
For parsing, we use ANTLR4~\cite{antlr4} and the example grammars provided by ANTLR\footnote{\url{https://github.com/antlr/grammars-v4}} for each of the languages in CodeSearchNet.
For each snippet, we attempt to parse it as a method (or the language equivalent thereof).
This non-standard entry point is needed as CodeSearchNet data points do not represent full compilation units in the ANTLR4 sense.
Not all snippets can be preprocessed in this manner, thus we filter any snippets that fail to parse. Further, PHP blocks are often processed into trivial ASTs that do not capture the syntax well due to a mix of PHP and HTML, thus we remove PHP from the data set.
\Cref{tab:results:corpora} shows the statistics of the resulting corpus.

To obtain vocabularies for each language, we consider the full data set; however, due to computational limits (VRAM, number of GPUs), we consider TreeLSTM models trained with sequences up to a certain sequence length ($k \in [20,35] \bigcup \{ 40, 50, 60\}$). 
Importantly for us, the languages that are more heavily filtered have more data points in total, hence we expect models to train on sufficiently many examples even after filtering.
Further, not all languages can be fully loaded for training at all sequence lengths (Python, Java, go), or there are insufficient training points for small sequences (Python --- from $28$, JavaScript --- from $24$), creating a censored view of the self-cross-entropy.
We, however, remark that there is a critical transition point between $k=20$ and $30$: Java at $21$, go at $22$, JavaScript at $26$, Ruby at $21$. We do not observe the transition for Python but hypothesise that it happens before $k=28$.

\subsection{Self-Cross Entropy of AST trees}
\label{sec:results:self}

\begin{figure}[!t]
    \centering
    \includegraphics[width=0.66\textwidth]{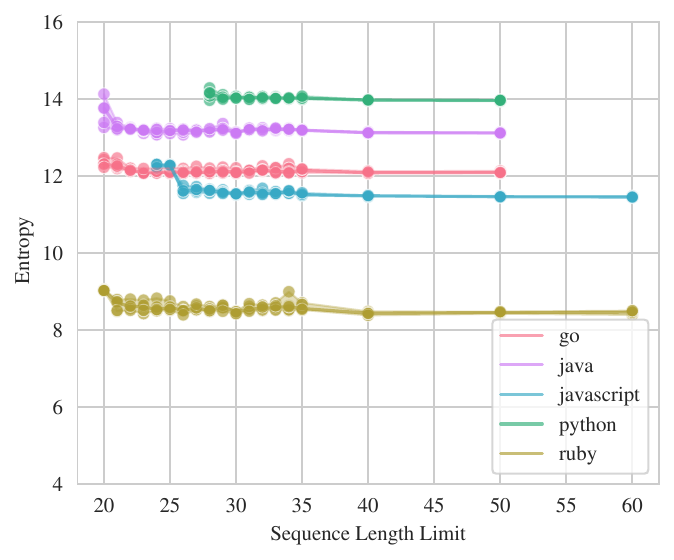}
    \caption{Self-cross-entropy of TreeLSTM for different limits of the sequence length across languages.}
    \label{fig:cross-ent}
\end{figure}

In this study, we replicated the self-cross-entropy setup from Hindle \etal~\cite{hindle2012naturalness} with a few augmentations as detailed in \Cref{sec:method}: we opted for five-fold cross-validation due to computational concerns, and instead of having an $n$ as in $n$-grams, we opted to control the sequence length as that serves as the closest proxy to the context size.

\Cref{fig:cross-ent} shows the self-cross-entropy for different languages and sequence length limits. 
While lacking in the granularity of data points compared to $n$-gram results before us, we still observe a shift towards lower entropy as we increase the limit, and thus context size. 
Languages tend to show this transition before $k = 30$.
While we lack data points for $k = 60$ for Java, go, and Python, as the transition happens at or before $k = 30$, we expect the final data point to be representative of these languages.
Compared to English under an $n$-gram model, whose self-cross-entropy is known to be around 12 bits, two languages show more predictability/lower self-entropy by being well under that, Go shows a borderline result by being a smidgen above 12 bits, while Java and Python show less predictability. This hints that the hypothesis should be checked before a model is employed with a particular structure as using an incorrect structural inductive bias/prior can misguide a model.

A new observation is a decrease in model self-cross-entropy variance as the context capacity increases. This suggests a transition from under-capacity where the model has to compress or learn to over-capacity where the model can start memorising data from the training set. We do not observe the modern interpolation regime which we expect to be at context sizes beyond those explored for this work.

A language that does not show this trend and instead shows a self-cross-entropy that is abnormally high is PHP which we removed from \Cref{fig:cross-ent}.
Upon manual inspection, we found that the ASTs produced by ANTLR4 in this case tended to produce shallow trees that quickly have a single HTML block into which most tokens were relegated.
This artificially loses information and makes PHP less predictable (under this view) than even Gutenberg Corpus English (which under a bi-gram model is ~12 bits).

\subsection{Comparison with Previous View}
\label{sec:results:comparison}

\begin{table}[t]
    \centering
    \caption[Comparison with $n$-gram]{The self-cross-entropy of an $n$-gram model and TreeLSTM model on ANTLRv4 ASTs after reaching ``steady-state''. Restricting to the languages studied by Rahman \etal~\cite{rahman2019natural} and those in \Cref{sec:results:self}, we find that $n$-gram captures the sequence level predictability better than a TreeLSTM model captures the tree level predictability.}
    \label{tab:results:comparative}
    \begin{tabular}{ l r r } 
    \hline 
    \textbf{Language} & {\textbf{$n$-gram~\cite{rahman2019natural}}} & {\textbf{TreeLSTM (ANTLRv4 AST)}}\\ 
    \hline
    Java & 8.0 & 13.1 \\
    JavaScript & 7.0 & 11.5 \\
    Python & 9.5 & 14.0 \\
    Ruby & 9.0 & 8.5 \\
    \hline 
\end{tabular}
\end{table}

Focusing on JavaScript, Java, Ruby, and Python which overlap with languages studied by Rahman \etal~\cite{rahman2019natural}, we find mixed results in terms of predictability; however, in line with the results of Rahman \etal after they account for SyntaxTokens.

For the sake of comparison, we will focus on the `steady state' results for both Rahman \etal and ourselves, that is, after any `state transition' due to sequence length for ourselves and results for 4-grams and higher for Rahman \etal.

What we observe is that Ruby is marginally better at 8.5 bits \vs 9 bits when viewed as a tree, while JavaScript suffers in predictability at 11.5 bits when viewed as a tree \vs 7 bits when viewed as an $n$-gram. For Java, we find that 13.1 \vs 8 bits, while Python 14.0 \vs 9.5 bits, hinting that the ASTs obtained from ANTLRv4 are a bad fit for the language model task explored. \Cref{tab:results:comparative} shows the results of this comparison.

The variability observed in Java for the results with a context size of $20$ is due to few training points and a high dependence on initial conditions; however, the trend seems to manifest even for Java despite this restriction to our experiment.

The Python result is anomalous, and we do not if the fall is before the observable data points (as would be hinted at by the other languages), or censored by the computational cost of running the experiment for longer sequences. Unlike PHP, the ASTs generated for Python are valid and diverse as determined by a manual inspection.

While a benefit of an AST view becomes the handling of SyntaxTokens ``for free'' by the structure; there can be a dependence on the grammar, parser, and other similar choices that can enable or hinder learning trees directly --- JavaScript and Ruby have demonstrated good performance, while Java and Python have not. This is despite all languages using the community-provided ANTLRv4 grammars for each respective language.

A similar de-correlation between AST predictions and masked-pre-trained language models is also observed by Velasco \etal~\cite{velasco2024masked4code}, which taken together with our results presented here, may suggest that the self-cross-entropy may measure the token or line-level prediction performance more than the syntactic performance that tree models target.
\section{From Naturalness to Predictions}
\label{sec:indirect}

To understand the hypothesis, we looked, thus far, to direct evidence by taking an approach similar to Hindle \etal~\cite{hindle2012naturalness} via self-cross-entropy showing mixed results. 
This is counter-intuitive to the success of the models, suggesting that perhaps they aid downstream tasks via a side channel rather than a more efficient encoding of the language itself.

In this section, we shift the focus to indirect, but more readily applicable evidence for the hypothesis. 
Previously, Ray \etal~\cite{ray2016naturalness} have shown that harder-to-predict lines tend to be more defect-prone.
Crucially for us, this suggests there is a statistical difference between defect prone and not that could be exploited by a model, which we explore at the start of \Cref{sec:indirect:jit}.
We study if exploiting such a signal can provide a model for JIT Defect Prediction which we demonstrate in \Cref{sec:indirect:results:jitdp}. 
We aim to show that a simple pipeline can get near state-of-the-art results without manual human feature engineering by exploiting the regularities we have demonstrated in \Cref{sec:indirect:jit} below.

\subsection{Exploiting Structure for Just-in-Time Defect Prediction}
\label{sec:indirect:jit}

\begin{figure}[!t]
\centering
\subfloat[Apache Commons Lang.]{\includegraphics[width=0.49\linewidth]{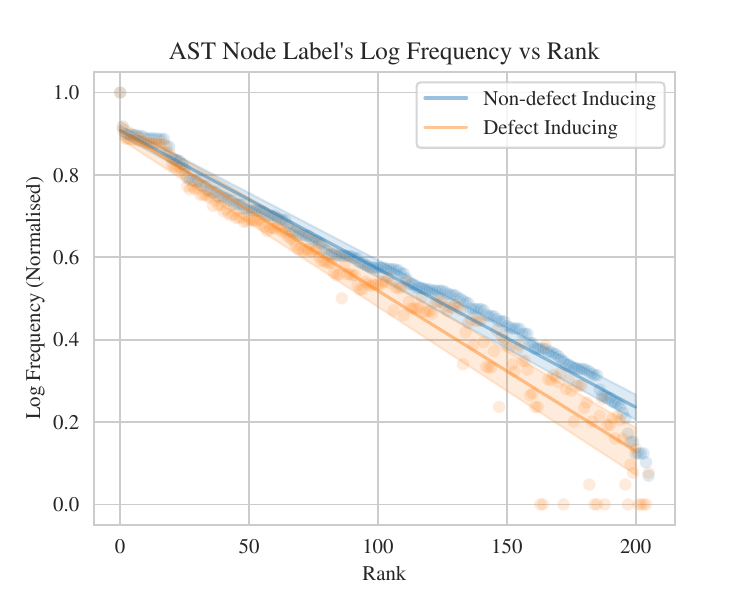}%
\label{fig:logfreq-v-rank:lang}}
\hfill
\centering
\subfloat[Apache Commons Math.]{\includegraphics[width=0.49\linewidth]{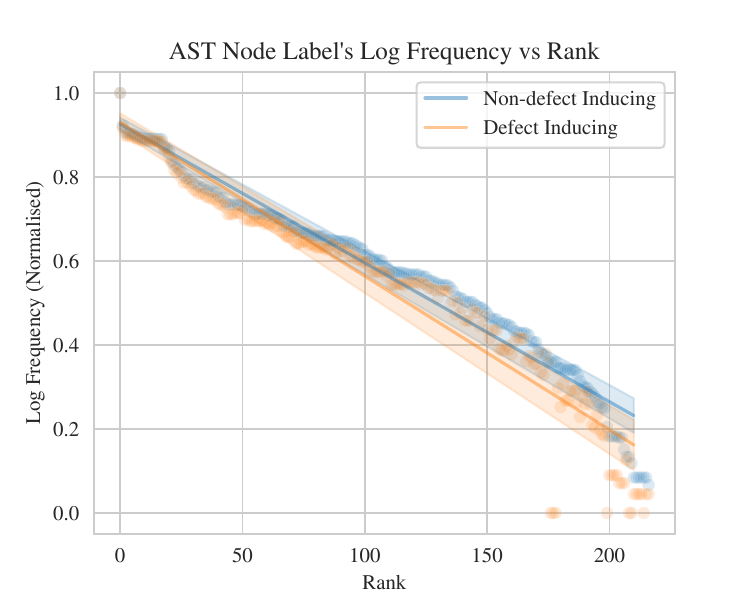}%
\label{fig:logfreq-v-rank:math}}
\hfill
\caption{Normalised log-frequency of AST node labels \vs their rank in Apache Commons Lang and Math. Hue shows if the commit is defect-inducing. The two samples (defect inducing and non-defect inducing) are of equal size as we use the same method before and after the introduction of the bug. 
}
\label{fig:logfreq-v-rank}
\end{figure}

Currently, in literature we see code structure being employed for more traditional language model tasks such as code completion~\cite{allamanis2017learning, bielik2016phog, raychev2016learning}, synthesis~\cite{allamanis2015bimodal, gvero2015synthesizing}, summarisation~\cite{fowkes2017autofolding, iyer2016summarizing}; however, Ray \etal show that naturalness can extend to fault localisation and defect prediction~\cite{ray2016naturalness}, and Khanfir \etal~\cite{khanfir2022codebert} show that a similar approach works using a transformer model at the line level.
Inspired by this, we devise an experiment to use structured naturalness, here manifested as the similarity between ASTs, for just-in-time (JIT) defect prediction, a more applied task than the self-cross-entropy experiment.

To motivate this approach, \Cref{fig:logfreq-v-rank} presents the log-frequency \vs rank correlation for AST node labels for two sub-populations from Apache Commons Lang: 
the same methods before and after the introduction of the defect --- hence by construction we have the same number of datapoints for both defective and good methods, avoiding any effect from different population sizes on the observed trends.
We can notice two main things: these labels exhibit a Zipfian distribution, as we already expected from the CodeSearchNet~\cite{Husain2019} results, and, when using the order deduced from the whole population of commits, defect-inducing commits exhibit a statistically different Zipfian distribution ($p<0.001$ per a Wilcoxon Signed-Rank test).
This leads to our approach which aims to test if the embedding space from the previous section can solve the classification task. Specifically, we use the embedding space from the Java TreeLSTM as a feature space for downstream classifiers, here Random Forests, to classify if a method is likely to induce defects.

In what follows, we first present the experimental set-up then briefly discuss the results on Apache Commons Lang and Apache Commons Math. 
This setup is limited in that our data set is two projects under the same Apache Commons umbrella; however, the results point towards promising directions.

\subsubsection{JIT Defect Prediction Experimental Set-up}
\label{sec:indirect:jit:setup}

To perform just-in-time defect prediction, we want to perform the following:
given a project version control (for example git repository), at each commit as (if it were) done by the developer, predict if this commit is likely to induce a new defect/bug or not.
This prediction set-up is in line with Sohn \etal~\cite{Sohn21fault}, \ie we employ longitudinal evaluation such that we only train on past data and not future commit information relative to a prediction target. 

Unlike Sohn \etal, we only make use of AST and commit diff data for our predictions, however, we do make use of the project issue tracker and the SZZ algorithm~\cite{sliwerski2005changes} to generate the same training data as Sohn \etal 
As we require AST data, we do not reuse the corpus directly from Sohn \etal instead doing a best-effort recreation of the corpus following the details from the paper.
Specifically, we make use of the collection period information from the Sohn \etal paper to filter the commits to be the same as theirs. 
For some commits, we fail to generate AST data within a one-hour time-out, causing minor divergences in the data set. 

\begin{table}[!t]
	\centering
	\caption{Corpus Statistics for Apache Commons Lang (top) / Math (bottom). The table presents the average number of classes touched by a commit, abstract syntax trees, and the number of defect-inducing commits in a fold and overall.}
	\label{tab:indirect:stats}
	\begin{minipage}{0.75\textwidth}
            \textbf{Apache Commons Lang}\\
	        {\begin{tabular}{ r S[table-format=1.2] S[table-format=1.2] S[table-format=5.0] S[table-format=4.0] S[table-format=3.0] S[table-format=3.0] }
\hline
\textbf{Fold} & \multicolumn{2}{c}{\textbf{Classes}} & \multicolumn{2}{c}{\textbf{ASTs}} & \multicolumn{2}{c}{\textbf{Defects}}\\
\hline
& \textbf{Train} & \textbf{Test} & \textbf{Train} & \textbf{Test} & \textbf{Train} & \textbf{Test} \\
\cline{2-7}
0 & 2.21 & 1.63 & 1008 & 744 & 18 & 45 \\
1 & 1.92 & 2.63 & 1752 & 1201 & 63 & 44 \\
2 & 2.16 & 2.22 & 2953 & 1013 & 107 & 62 \\
3 & 2.17 & 2.38 & 3966 & 1087 & 169 & 108 \\
4 & 2.22 & 2.32 & 5053 & 1059 & 277 & 122 \\
\hline
\textbf{O/A} & 2.13 & 2.24 & 14732 & 5104 & 634 & 381
\end{tabular}}
	\end{minipage}
    \\
    \vspace{1em}
    \begin{minipage}{0.75\textwidth}
            \textbf{Apache Commons Math}\\
            {\begin{tabular}{ r S[table-format=1.2] S[table-format=1.2] S[table-format=5.0] S[table-format=5.0] S[table-format=4.0] S[table-format=3.0] }
\hline
\textbf{Fold} & \multicolumn{2}{c}{\textbf{Classes}} & \multicolumn{2}{c}{\textbf{ASTs}} & \multicolumn{2}{c}{\textbf{Defects}}\\
\hline
& \textbf{Train} & \textbf{Test} & \textbf{Train} & \textbf{Test} & \textbf{Train} & \textbf{Test} \\
\cline{2-7}
0 & 5.16 & 6.31 & 4500 & 5480 & 22 & 148 \\
1 & 5.74 & 2.61 & 9980 & 2263 & 170 & 76 \\
2 & 4.69 & 2.74 & 12243 & 2380 & 246 & 72 \\
3 & 4.21 & 3.51 & 14623 & 3051 & 318 & 228 \\
4 & 4.07 & 4.43 & 17674 & 3844 & 546 & 231 \\
\hline
\textbf{O/A} & 4.77 & 3.92 & 59020 & 17018 & 1302 & 755
\end{tabular}}
    \end{minipage}
\end{table}

The final AST corpora generated are as follows:
for Apache Commons Lang, we obtain $14732$ training ASTs, from which $634$ are defect-inducing, and we have $5104$ test ASTs, among which $381$ are defect-inducing;
for Apache Commons Math --- $59020$ training ASTs, from which $1302$ are defect-inducing, while the test set has $17018$ ASTs with $755$ defect-inducing.

Per fold, these vary from the smallest first prefix at $1008\ (18)$~/ $4500\ (22)$ training and $744\ (45)$~/ $2263\ (72)$ test ASTs, to the largest last prefix --- $5053\ (277)$~/ $17647\ (546)$ training and $1059\ (122)$~/ $5480\ (148)$ test ASTs shown as ``total (defect inducing)'' where the first pair is for Lang and the second for Math. 
\Cref{tab:indirect:stats} shows these numbers.

\begin{figure}[!t]
    \centering
    % trim is <left> <lower> <right> <upper>
    \includegraphics[width=0.99\textwidth, trim=0 0 0 0]{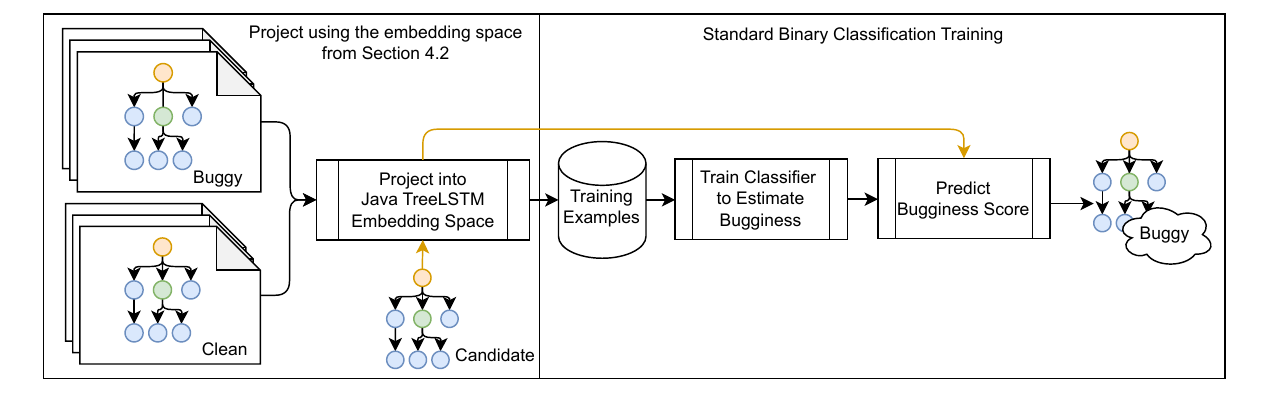}
    \caption{The training process for determining if a method AST is defect-inducing or clean. We assume that the input has already been preprocessed to be ASTs. The orange lines show how a candidate AST passes through the system, \ie is transformed by the embedding space projection before being classified by a Random Forest Classifier.}
    \label{fig:cls_training}
\end{figure}

In our implementation, 
we first determine affected ASTs by finding the before/after method source code for each method edited.
We then convert our ASTs to a vector space using the learnt embedding space for Java for sequences of length 30 (where we observe the second dip in entropy in \Cref{fig:cross-ent}). 
Thus, we use the features learnt under masked language pre-training by the TreeLSTM model, obtaining one vector per method-level AST.

To predict if a commit is defect inducing, we train a Random Forest Classifier (with the default hyper-parameters of the scikit-learn implementation~\cite{SciPy} 
except for the number of trees which we set to 200 to match Sohn \etal)
on our feature vectors, and consider a commit to be defect-inducing if any method is determined to be defect-inducing, \ie a single defect-inducing method ``taints'' the whole commit.
\Cref{fig:cls_training} shows the training process employed.
As our aim is to explore the viability of using a simple method to extract information from the embedding space, we do not explore the hyperparameter space or the choice of classifier, opting for a standard choice within SE literature instead.

For evaluation and to better capture the temporal nature of this problem, we perform longitudinal evaluation and make use of the scikit-learn TimeSeriesSplit\footnote{\url{https://scikit-learn.org/stable/modules/generated/sklearn.model_selection.TimeSeriesSplit.html}} method~\cite{SciPy}.
We split our data into 5 prefixes and always evaluate the next 10\% of data chronologically. 
Due to development activity, this creates a tough first fold for Apache Commons Math as the first row on the right side of \Cref{tab:indirect:stats} shows.
This longitudinal evaluation ensures we do not use commit or project information that occurs after a prediction (future data) when evaluating the correctness of that prediction.
We use a 5-fold longitudinal split to match the protocol from Sohn \etal, thus differing only on the suffix being 10\% rather than a fixed 30,60,90,120-day window.
This last choice simplifies the experimental protocol.

\subsubsection{Just-in-Time Defect Prediction Results}
\label{sec:indirect:results:jitdp}

\begin{table}[!t]
	\centering
	\caption{Performance metrics for Apache Commons Lang (top) / Math (bottom) per fold and overall for a Random Forest Classifier using the Java embedding space from \Cref{sec:results:self}. `US' denotes if we use under-sampling.}
	\label{tab:indirect:results:rf}
 	\begin{minipage}{0.75\textwidth}
            \textbf{Apache Commons Lang}\\
	        {\begin{tabular}{l c S[table-format=-1.3] S[table-format=-1.3] S[table-format=-1.3] S[table-format=-1.3] S[table-format=-1.3] }
    \hline
    \textbf{Fold} & \textbf{US} & \textbf{Acc} & \textbf{P} & \textbf{R} & \textbf{F1} & \textbf{MCC}\\
    \hline
    0 & $\times$ & 0.523 & 0.250 & 0.068 & 0.107 & 0.085 \\
    & \checkmark & 0.586 & 0.189 & 0.318 & 0.237 & 0.138 \\
    1 & $\times$ & 0.535 & 0.182 & 0.136 & 0.156 & 0.081 \\
    & \checkmark & 0.546 & 0.123 & 0.386 & 0.187 & 0.059 \\
    2 & $\times$ & 0.539 & 0.217 & 0.169 & 0.190 & 0.088 \\
    & \checkmark & 0.575 & 0.174 & 0.508 & 0.259 & 0.104 \\
    3 & $\times$ & 0.533 & 0.340 & 0.167 & 0.223 & 0.087 \\
    & \checkmark & 0.580 & 0.323 & 0.463 & 0.380 & 0.144 \\
    4 & $\times$ & 0.651 & 0.465 & 0.532 & 0.496 & 0.291 \\
    & \checkmark & 0.607 & 0.336 & 0.831 & 0.478 & 0.204 \\
    \hline
    \textbf{O/A} & $\times$ & 0.556 & 0.291 & 0.216 & 0.236 & 0.198 \\
    & \checkmark & 0.578 & 0.228 & 0.501 & 0.308 & 0.178 \\
\end{tabular}}
	\end{minipage}
    \\
    \vspace{1em}
    \begin{minipage}{0.75\textwidth}
            \textbf{Apache Commons Math}\\
            {\begin{tabular}{l c S[table-format=-1.3] S[table-format=-1.3] S[table-format=-1.3] S[table-format=-1.3] S[table-format=-1.3] }
    \hline
    \textbf{Fold} & \textbf{US} & \textbf{Acc} & \textbf{P} & \textbf{R} & \textbf{F1} & \textbf{MCC}\\
    \hline
    0 & $\times$ & 0.548 & 0.328 & 0.164 & 0.219 & 0.131 \\
    & \checkmark & 0.582 & 0.277 & 0.342 & 0.306 & 0.152 \\
    1 & $\times$ & 0.611 & 0.158 & 0.481 & 0.238 & 0.143 \\
    & \checkmark & 0.638 & 0.134 & 0.802 & 0.230 & 0.161 \\
    2 & $\times$ & 0.581 & 0.144 & 0.384 & 0.209 & 0.108 \\
    & \checkmark & 0.607 & 0.121 & 0.730 & 0.208 & 0.122 \\
    3 & $\times$ & 0.632 & 0.490 & 0.432 & 0.459 & 0.277 \\
    & \checkmark & 0.655 & 0.376 & 0.806 & 0.513 & 0.279 \\
    4 & $\times$ & 0.641 & 0.412 & 0.614 & 0.493 & 0.257 \\
    & \checkmark & 0.593 & 0.325 & 0.871 & 0.473 & 0.189 \\
    \hline
    \textbf{O/A} & $\times$ & 0.603 & 0.306 & 0.415 & 0.324 & 0.203 \\
    & \checkmark & 0.615 & 0.246 & 0.710 & 0.346 & 0.196 \\
\end{tabular}}
    \end{minipage}
\end{table}

Our results are restricted to Apache Commons Lang and Apache Commons Math projects due to the computational costs of generating ASTs for all methods touched by the history of a project.
\Cref{tab:indirect:results:rf} shows the per fold and overall results. Folds can be interpreted as increasing training data as the training prefix increases steadily from 20\% of the data on fold 0 up to 80\% on fold 4.
The best reported F1 by Sohn \etal on Apache Commons Lang is $0.199$ while our approach reports $0.236$.
While balanced accuracy is $0.556$ for us and $0.596$ for Sohn \etal
On Apache Commons Math, we find that the F1 is $0.401$ for Sohn \etal while we obtain $0.324$, a worse result when considered directly; accuracy is $0.603$ for us v $0.690$ for Sohn \etal, both results close to SOTA. 
The results we observe are in line with what one can intuit from \Cref{fig:logfreq-v-rank}.
Indeed, there is a more significant overlap between the tail of the defect-inducing and non-defect-inducing commit distributions for Apache Math compared to Apache Lang.
This suggests that the classification task, as formulated for our approach, is more difficult for the former.

We have to remark, however, that evaluation protocols differ between the papers so the numbers should not be interpreted as a direct comparison, rather this suggests that the simplified approach presented here holds promise for JIT defect prediction.
Notably, we evaluate on a 10\% of all data suffix while Sohn \etal employ a more realistic 30, 60, 90, and 120 days suffix.
We made our evaluation protocol choice to maintain the simplicity of a proof-of-concept experiment that makes use of only naturalness at a structural level.

As defect-inducing commits are rare, our training data is imbalanced and performance can be further improved by employing edit nearest neighbours (ENN) under-sampling~\cite{hattori2000new}.
We explore this direction as well and observe the following main takeaways:
Accuracy increases marginally from $0.556$ to $0.578$ (respectively $0.603$ to $0.615$ for Math), while precision falls from $0.291$ to $0.228$ (respectively $0.306$ to $0.246$).
Recall, however, increases from $0.216$ to $0.501$ which in turn increases F1 from $0.236$ to $0.308$ improving over Sohn \etal (respectively, recall --- from $0.415$ to $0.710$, and F1 --- $0.324$ to $0.346$ approaching but not reaching the Sohn \etal result). 
MCC remains relatively unchanged: $0.198$ v $0.178$ (respectively --- $0.203$ v $0.196$).
\Cref{tab:indirect:results:rf} presents these results per fold and overall where the `US' column has a `\checkmark'.

Qualitatively, we also observe the model trained with under-sampled data to become generally more likely to suggest that a commit is defect-inducing.
Ideally, such a tool should be high precision to not bother developers needlessly; however, due to the precision with and without under-sampling being too low for such use-cases, trading precision for recall can be worthwhile since this leads to an increase in F1 score.

These results show that our approach guided by intuition from structured naturalness enables a competitive and simple approach to just-in-time defect prediction; 
\emph{crucially, we avoid manually constructing a feature space for the problem and instead exploit statistical properties of structures from the populations of interest.}
The design space, however, is still open for investigation, either from a better understanding of structured naturalness or combining more traditional features employed in defect prediction literature.

\section{Threats to Validity}
\label{sec:threats}

Our study faces the usual threats to its external validity: the degree to which our findings generalise depends on how representative CodeSearchNet~\cite{Husain2019} is as a data set.
We consider this risk to be mitigated by CodeSearchNet being a large sample from GitHub and thus should be representative of at least OSS projects.

Our preprocessing using ANTLR4 and their public grammars introduces additional bias in which snippets fail to parse or parse in a trivial and uninformative way.
Our results for PHP are an example of this latter.
Due to the chosen entry point for PHP producing trivial ASTs, we have discarded the data from the reported results as they do not represent models on proper ASTs.
Still, snippet parsing in general is a challenging problem and we consider the amount of data generated sufficient for the languages we manage to report results on.

Another bias is introduced by our filtering by sequence length.
We do not study the (un-)naturalness of large ASTs, to a large degree, by design.
We do such filtering to enable a proxy for the context window of a $n$-gram model that is reasonable for the TreeLSTM model and for computational reasons --- performing five-fold cross-validation for TreeLSTM models on certain CodeSearchNet languages, such as Java, Python, and Go, that have large vocabularies require significant compute.
However, this decision can hide issues due to long-range dependencies as these may not exist even in our largest sequences.
Solving such a long-range dependency issue would require a transformer model with a sufficiently large attention window rather than an LSTM adapted to trees which is beyond the scope of this research.
The aim of this work is not to produce state-of-the-art language models or large language models, rather, we aim to provide empiric evidence that models that use structural information capture similar levels of predictability as a token level while overcoming issues highlighted by previous work~\cite{rahman2019natural} regarding trivial predictability, or, when they fail to do so such as for Java, they provide a signal side-channel that can still help downstream tasks, demonstrated through JIT defect prediction.

A threat to our conclusion is in our choice of baseline.
We compare with Sohn \etal~\cite{Sohn21fault} on defect prediction rather than with a transformer-based model such as CodeBERT~\cite{khanfir2022codebert} tuned to predict line-level bugginess.
Our choice of baseline is due to working on snippets rather than line-granular and the mapping between the two in our case is non-trivial.
A separate threat to the conclusion is our choice of a tree-based model. We chose a Child-Sum TreeLSTM model~\cite{tai2015improved}, rather than a TreeTransfomer~\cite{wang2019tree} or an N-way TreeLSTM. This does mean that our model loses information about the order of siblings (as the sum aggregation masks this information) and may not be fully ideal for the language modelling task. 
However, the result on the downstream task indicates that despite such shortcomings, tasks can exploit the statistical differences due to naturalness for defect prediction.

Finally, we look at self-cross-entropy as a measure of predictability or naturalness. 
We do not explore a diverse range of downstream tasks for such tree-based representations.
To partially alleviate this concern, we perform a small scale just-in-time defect prediction task by using the embedding space for Java from the language modelling pretraining instead of hand-engineered features to reach near-SOTA on the task with minimal manual effort.

\section{Related Work}
\label{sec:rel_work}

\emph{Naturalness and language models:} 
From the original work on the naturalness of source code by Hindle \etal~\cite{hindle2012naturalness}, a plethora of work has followed employing language models for software-engineering tasks, initially for code completion in the Hindle \etal paper as well as Bielik \etal~\cite{bielik2016phog} and Raychev \etal~\cite{raychev2016learning} among many. 
This was followed by other tasks that can benefit from a language model such as code synthesis~\cite{allamanis2015bimodal, gvero2015synthesizing}.
Most of the work focused on the predictability at the token level employing $n$-gram models~\cite{ray2016naturalness}, RNNs~\cite{dam2016deep}, or CNNs~\cite{allamanis2016convolutional}, in short, approaches that can best exploit the sequence nature of source-code. 
Another line of work veered towards ``unnatural'' code. 
One strand of work looked at detecting syntax errors using naturalness~\cite{campbellSyntaxErrors2014}, while another took an outlier detection approach towards fault localisation~\cite{ray2016naturalness}. 
With the rise of transformer-based models, Khanfir \etal have measured the naturalness of code as viewed by CodeBERT~\cite{khanfir2022codebert}.
Khanfir \etal's methodology follows that of Ray \etal~\cite{ray2016naturalness} and they focus on detecting unnatural code as buggy rather than directly measuring self-cross-entropy as done by Hindle \etal

These lines of work do not consider syntax parses or other graph data that can help focus the signal they wish to exploit. 
The shape of such a signal is crucial.
Indeed, picking the appropriate statistical tools for a task requires knowledge of the regularities in ASTs and PDGs as induced by those at the token level.

A recent line of work focused on large language models (LLMs) and their capacity to generalise as few- or, indeed, zero-shot learners and fine-tuning them for source code. 
The former focuses on In-context Learning~\cite{lampinen2022can, brown2020language} on foundational models such as GPT-3.5~\cite{openai2022gpt3.5}, GPT-4~\cite{openai2023gpt4}, GPT-4o~\cite{openai2024gpt4o}, or Llama~\cite{touvron2023llama, llama3modelcard}, while the latter, focuses on producing source-code specific models such as StarCoder(v2)~\cite{li2023starcoder, lozhkov2024starcoder}. 
While they do not explicitly learn source-code embedding as a primary task, they serve as a basis for the same downstream tasks of source-code generation, summarisation, tab completion \etc, and more recently even compiler optimisation~\cite{cummins2023large}.
Building upon this work, and borrowing from the recent prompt engineering literature, Ahmed \etal focus on exposing code information to an LLM via the prompt~\cite{ahmed2023improving}.
In ML literature, work has focused on embedding graph knowledge into LLMs for query~\cite{yasunaga2021qa, jiang2023structgpt}. To our knowledge, no work has combined the two aspects to provide LLMs with direct access to dataflow graphs or ASTs.
Measuring the self-cross-entropy of LLMs is complicated by their exposure to extensive datasets, usually subsuming those employed in our work here, as well as their computational cost for inference making cross-validation prohibitive, therefore, they are outside the scope of this work.

\emph{Use of structure for predictions:} 
More recent work has started incorporating AST or graph data into their models.
Most work using such a signal focuses on linearising the tree data and considers AST paths (leaf-to-leaf, root-to-leaf) as an additional sequence signal.
code2seq~\cite{alon2018code2seq} uses such paths through an AST together with an attention mechanism to better model source code. 
They demonstrate the usefulness of their model on code summarisation and code captioning.
To use tree information more directly, Dam \etal employ a TreeLSTM for software defect prediction, augmenting the LSTM model to architecturally consider the tree structure of the data~\cite{dam2018deep}.
Fernandes \etal~\cite{fernandes2018structured} augment natural language sentences with parse information to enable information to propagate across longer ranges. 
They do so by augmenting existing seq2seq methods with a graph structure and a GNN to propagate information. 
Allamanis \etal~\cite{allamanis2018survey} provide a more detailed view of both token level and structured models in literature in their review.
Kovalenko \etal~\cite{Kovalenko2019pathminer} create a tool to facilitate mining paths from syntax trees and provide a common basis for other researchers to build upon for use in downstream Machine Learning tasks. Here, paths can be leaf-to-leaf, root-to-leaf (root-paths) \etc These provide a common abstraction similar to $n$-gram or bag of words over sequence data.

\emph{Dual channel theory:} 
A recent line of work focuses on the idea that code manifests itself across two channels, a natural language channel (such as comments, identifier names \etc) which is aimed at developers and a formal, algorithmic channel aimed both for machine execution and developer understanding.
Casalnuovo \etal~\cite{casalnuovo2020theory} present this theory and argue the benefits of research that considers both channels can bring, for example by focusing on defect localisation techniques where the channels are not in sync.
Work that uses this framework predating the introduction of the theory exists, Dash \etal~\cite{refinym2018dash} use this meta-framework for identifier name prediction by considering identifiers together with data-flow, creating name-flows.

The structured naturalness hypothesis is orthogonal to this research line; however, with potential mutual benefits.
By considering the structure of both channels, better constraints can be formulated enabling a better study of how the channels synchronise, de-synchronise, or indeed how to spot issues and re-synchronise them.

\emph{Just-in-time defect prediction:} 
Sohn \etal~\cite{Sohn21fault} employ fault localisation to improve defect prediction to make it more directly actionable and accurate. 
Their work differs from previous literature which focused mostly on exploiting correlations between past and future defects~\cite{graves2000predicting, hassan2005top, zimmermann2007predicting}.
Our application of the structured naturalness hypothesis follows the spirit of that in Sohn \etal; however, we focus only on using AST features to better align the task with the hypothesis we are exploring.
\section{Conclusion}
\label{sec:conclusion}

In this paper, we formulate the structured naturalness hypothesis: code, when viewed in a structured fashion, be the structure syntactic or semantic, will exhibit rich get richer effects over substructures present. 
Indeed, current literature either uses paths through structures to produce sequences or encodes the structure directly into model architecture. 
The former is a sign of a mismatch between statistical approaches and the nature of the data researchers want to use.
Further, performance can be impacted by the choice of how to project structure into sequences, which ideally should not matter.
The latter approach, encoding the code structure into a neural architecture, leaves us dissatisfied due to the black-box nature of neural networks and can have high data costs associated with their use.
While our study is limited to tree models: we use only a TreeLSTM model to gather empirical evidence for the hypothesis, we expect our results to extend to graphs and other tree models.
In particular, we found that even when a structure is more difficult to model --- Java was less predictable under a Child-Sum TreeLSMT --- when employed for a downstream task --- JIT defect prediction --- there was sufficient signal both in terms of the Zipf distributions of the two populations as well as for a simple classification pipeline using the TreeLSTM Java embedding space to learn to predict defect proneness of commits.
This latter result is particularly interesting as it suggests that the model learns features useful for downstream tasks that are not immediately measured by the self-cross-entropy measure from Hindle \etal~\cite{hindle2012naturalness}.

While extending the hypothesis to natural language (NL) is more ambitious compared to programming languages (PL), similar structured views can be formed for natural language as well: be that parse trees, dependency parses, entity graphs \etc
Existing literature in Natural Language Processing makes use of Tree Kernels~\cite{Moschitti2008treekernel} for semantic labelling to make the task amenable to perceptron or support vector machine approaches. This view has not been extended for other tasks nor has the regularities of such structured views been studied as a main subject.
We expect to some degree our hypothesis to extend beyond just code: akin to how the Zipfian nature crossed the boundary from NL to PL, we expected the structured naturalness hypothesis to cross the boundary back.

By improving the tooling available on both sides: NL and PL, we also hope this can extend to cross-channel constraints as explained by Casalnuovo \etal~\cite{casalnuovo2020theory}. For example, studying the structures that arise in both natural and programming languages in literate programming~\cite{knuth1984literate} can yield insight into how structures across them can align to reinforce or explain each other. 

In this current work, we looked towards evidence in the case of trees towards the structured naturalness hypothesis which can pose a conclusion threat to validity. 
Hence, immediate future work will focus on studying the structured naturalness hypothesis directly over graph structures as well, in terms of statistical regularities in code structures. 

We hope to spark interest in the domain of structured naturalness.
The tooling employed in our research should not be just pragmatic, rather it should help us glean insights that under a different view or framework would not be visible.
By putting a magnifying glass towards the structure of code, we hope to provide insight into code-at-scale that exists as a statistical phenomenon due to our collective development efforts.

\paragraph{Acknowledgements:} This work was supported by JST, CREST Grant Number JPMJCR22D3, Japan, and JSPS KAKENHI Grant Number JP21H03503.

\bibliographystyle{unsrtnat}
\bibliography{structnat}
\end{document}